\documentclass[11pt]{amsart}
\usepackage{amsfonts,amsmath,amssymb}
\usepackage[all]{xy}
\begin{document}

\newcommand{\fa}{{\mathfrak a}}
\newcommand{\Alg}{{\rm Alg}}
\newcommand{\ta}{{\tilde a}}
\newcommand{\BAlg}{{\bf{A}{\rm lg}}}
\newcommand{\Bm}{{\boldsymbol{\mathfrak I}}}
\newcommand{\bm}{{\mathfrak I}}
\newcommand{\bc}{{\partial}}
\newcommand{\fA}{{\mathfrak A}}
\newcommand{\fF}{{\mathfrak F}}
\newcommand{\C}{{\mathbb C}}
\newcommand{\R}{{\mathbb R}}
\newcommand{\Hom}{{\rm Hom}}
\newcommand{\Emb}{{\rm Emb}}
\newcommand{\HEmb}{{\bf{E}{\rm mb}}}
\newcommand{\HHom}{{{\bf H}{\rm om}}}
\newcommand{\mor}{{\rm mor}}
\newcommand{\out}{{\rm Out}}
\newcommand{\diff}{{\rm Diff}}
\newcommand{\PGl}{{\rm PGl}}
\newcommand{\xa}{{\bf{a}}}
\newcommand{\xb}{{\bf{b}}}
\newcommand{\xc}{{\bf{c}}}
\newcommand{\xd}{{\bf{d}}}
\newcommand{\xI}{{\bf{I}}}

\title{Quantized timelines}
\author{Jack Morava}
\address{The Johns Hopkins University,
Baltimore, Maryland 21218}
\email{jack@math.jhu.edu}
\date{21 May 2012}
\begin{abstract}{Conformal nets are a classical [5] topic in quantum
field theory: they assign operator algebras to one-dimensional manifolds,
and have close connections with one-dimensional topological field theories [1,4,7].
\medskip

\noindent
It seems to be well-known that the usual axioms for these constructions 
imply close relations between the action of the projective group on the 
line, and Connes' intrinsic flow on $C^*$-algebras. This note attempts to pin
down this specific fact, in terms of a category of (noncommutative) algebras 
and equivalence classes, under inner automorphisms, of homomorphisms between 
them. That category may be of independent interest.}\end{abstract}

\maketitle

\section{ A two-category of algebras} \bigskip

\noindent
{\bf 1.0} In the following, $A,B,C\dots$ will be $k$-algebras, eg for
$k = \C$, and $\Hom(A,B)$ will denote the set of algebra homomorphisms between
them. 
\[
\sigma_a(x) = a^{-1}xa := x^a
\]
will denote conjugation by a unit $a$ in some algebra (so $\sigma_{ab} = \sigma_b \circ \sigma_a$).
In fact the algebras of most interest here will be topological, and the homomorphisms will 
usually be continuous, but I'll leave that in the background. \bigskip

\noindent
Let $\HHom(A,B)$ be the category with $\Hom(A,B)$ as its set of objects, and morphisms
$(a,b): \phi_0 \to \phi_1$ defined by
\[
\mor(\phi_0,\phi_1) := \{(a,b) \in A^\times \times B^\times \; | \; \sigma_b \circ \phi_0 = \phi_1
\circ \sigma_a \}
\]
ie $\phi_1(x) \cdot \phi_1(a)b^{-1} = \phi_1(a)b^{-1} \cdot \phi_0(x)$ or, alternately,
\[
\sigma_{\phi_1(a)b^{-1}}(\phi_1(x)) = \phi_0(x) \;.
\]
Note, for future reference, that $\phi_1(a)b^{-1} = b^{-1}\phi_0(a)$ : take $x=a$ in the defining
condition for a morphism. \bigskip

\noindent
Morphisms $(a_0,b_0) \in \mor(\phi_0,\phi_1)$ and $(a_1,b_1) \in
\mor(\phi_1,\phi_2)$ compose according to the diagram
\[
\xymatrix{
A \ar[d]^{\phi_0}  \ar[r]^{a_0} & A \ar[d]^{\phi_1} \ar[r]^{a_1} & A \ar[d]^{\phi_2} \\
B \ar[r]^{b_0} & B \ar[r]^{b_1} & B }
\]
ie 
\[
\;(a_1,b_1) \circ (a_0,b_0) := (a_0a_1,b_0b_1) \in \mor(\phi_0,\phi_2) \;.
\]\bigskip

\noindent
$\HHom(A,B)$ is thus a groupoid. For example, the group of automorphisms of $\phi$ consists of 
pairs $a \in A^\times, b \in B^\times$ such that $\phi(a)b^{-1}$ lies in the center of the 
image of $\phi$: thus 
\[
\phi(\alpha)\beta^{-1} \cdot \phi(a) b^{-1} = \phi(a) \phi(\alpha)\beta^{-1} b^{-1} = \phi(a
\alpha)(b \beta)^{-1} \;.
\] \bigskip

\noindent
{\bf 1.1 Proposition:} The composition law 
\[
(\phi,\psi) \mapsto \psi \circ \phi : \HHom(A,B) \times \HHom(B,C) \to \HHom(A,C)
\]
defined on morphisms by 
\[
(a,b_0) \times (b_1,c) \mapsto (a,c \psi_1(b_1^{-1}b_0)): \psi_0 \circ \phi_0 \mapsto \psi_1 \circ \phi_1 
\]
defines a two-category $\BAlg_\out$ (of algebras and homomorphisms, up to inner automorphism).\bigskip

\noindent
{\bf Proof:} This composition law is well-defined on morphisms: we have
\[
\phi_1(a)b_0^{-1} \cdot \phi_0(x) = \phi_1(x) \cdot \phi_1(a)b_0^{-1}
\]
and
\[
\psi_1(b_1)c^{-1} \cdot \psi_0(y) = \psi_1(y) \cdot \psi_1(b_1)c^{-1}
\]\bigskip

\noindent
($\forall x,y \in A,B$, with $a \in A, \; b_0,b_1 \in B, c \in C$), so if we apply 
$\psi_0$ to the first expression, and left-multiply both sides by $\psi_1(b_1)c^{-1}$, we
get
\[
\psi_1(b_1)c^{-1} \cdot \psi_0(\phi_1(a)b_0^{-1}) \cdot (\psi_0 \phi_0)(x) = \psi_1(b_1)c^{-1}
\cdot \psi_0(\phi_1(x)) \cdot \psi_0(\phi_1(a)b_0^{-1}) \;,
\]
which in turn equals
\[
\psi_1(\phi_1(x)) \cdot \psi_1(b_1)c^{-1} \cdot \psi_0(\phi_1(a)b_0^{-1}) \;;
\]
that is, conjugation by $\psi_1(b_1)c^{-1}\psi_0(\phi_1(b_1)b_0^{-1})$ sends $\psi_1(\phi_1(x))$
to $\psi_0(\phi_0(x))$.\bigskip  

\noindent
On the other hand, it follows from the second relation above that
\[
\psi_1(b_1)c^{-1} \cdot \psi_0(\phi_1(b_1)b_0^{-1}) = \psi_1(\phi_1(a)b_0^{-1})\cdot \psi_1(b_1)c^{-1} 
= \psi_1(\phi_1(a)b_0^{-1}b_1)c^{-1} \;,
\]
so
\[
\sigma_{(\psi_1 \circ \phi_1)(a)(c\psi_1(b_1^{-1}b_0))^{-1}}((\psi_1 \circ \phi_1)(x)) = (\psi_0 \circ \phi_0)(x)
\]
ie $(a,c\psi_1(b_1^{-1}b_0)$ defines a morphism from $\psi_0 \circ \phi_0$ to $\psi_1 \circ \phi_1$. \bigskip

\noindent
It remains to check associativity. Let $\phi,\psi,(a,b_0),(b_1,c)$ be as above, and let 
\[
(z,\ta) \in \mor(\theta_0,\theta_1)
\]
for morphisms $\theta_0, \theta_1 \in \Hom(Z,A)$; then
\[
(z,\ta) \times (a,b_0) = (z,b_0 \phi_1(a^{-1}\ta)) : \theta_0 \times \phi_0 \to \theta_1 \times \phi_1 \;,
\]
so 
\[
((z,\ta) \times (a,b_0)) \times (b_1,c) = (z,c \psi_1(b_1^{-1}b_0 \phi_1(a^{-1}\ta))) :
(\theta_0 \times \phi_0) \times \psi_0 \to (\theta_1 \times \phi_1) \times \psi_1 \;;
\]
while
\[
(a,b) \times (b_1,c) \mapsto (a,c \psi_1(b_1^{-1}b_0)) : \phi_0 \times \psi_0 \to  \phi_1 \times \psi_1  \;,
\]
so 
\[
(z,,\ta) \times ((a,b)) \times (b_1,c)) = (z,c \psi_1(b_1^{-1}b_0)(\psi_1 \phi_1)(a^{-1}\ta)) \;.
\] 
$\Box$ \bigskip

\noindent
{\bf 1.2 Definition} By analogy with the construction of the group $\out(G) = {\rm Aut}(G)/(G/Z(G))$ of outer automorphisms 
of a group $G$, let
\[
\Hom_\out(A,B) := \pi_0 \HHom(A,B)
\]
be the set of isomorphism classes of objects in the groupoid $\HHom(A,B)$; this defines a category 
$\Alg_\out$ whose objects are algebras, and whose morphisms are equivalence classes, up to inner 
automorphism, of algebra homomorphisms. \bigskip

\noindent
{\bf Note} that a more familiar two-category of associative algebras takes bimodules as its
morphisms.\bigskip

\section {A two-category of one-manifolds} \bigskip

\noindent
{\bf 2.1} Let $I,J,\dots$ be compact connected oriented Riemannian one-manifolds (roughly: `intervals'), and 
let $\HEmb(I,J)$ be the category with oriented smooth embeddings 
\[
\epsilon: I \to J \in \Emb(I,J)
\]
as objects, and diagrams
\[
\xymatrix{
I \ar[d]^a \ar[r]^{\epsilon_0} & J \ar[d]^b \\
I \ar[r]^{\epsilon_1} & J }
\]
as morphisms $(a,b) : \epsilon_0 \to \epsilon_1$: where $\epsilon_0,\epsilon_1$ are embeddings as above, 
and $a,b$ are (orientation-preserving) diffeomorphisms supported in the {\bf interior} of the interval 
(ie each equals the identity in a neighborhood of the boundary of its domain). \bigskip

\noindent
{\bf 2.2 Lemma:} If $\epsilon \in \Emb(I,J)$ and $c \in \diff_0(I)$ is a diffeomorphism 
of $I$ supported, as above, in the interior of $I$, then 
\[
c^\epsilon(t) := (\epsilon \circ c)(\epsilon^{-1}(t)) \; {\rm if} \; t \in {\rm image} \; \epsilon
\]
(and $=t$ otherwise) defines an element of $\diff_0(J)$ such that 
\[
\xymatrix{
I \ar[d]^c \ar[r]^\epsilon & J \ar@{.>}[d]^{c^\epsilon} \\
I \ar[r]^\epsilon & J }
\]
commutes. $\; \; \; \; \; \Box$ \bigskip

\noindent
{\bf 2.3 Proposition:} The composition law
\[
(\epsilon,\delta) \mapsto \delta \circ \epsilon : \HEmb(I,J)\times \HEmb(J,K) \to \HEmb(I,K)
\]
defined on morphisms by 
\[
(a,b_0) \times (b_1,c) \mapsto (a, c(b_1^{-1}b_0)^{\delta_0}) 
\]
defines a two-category $\Bm_\bc$ (of intervals with collared boundaries). \bigskip

\noindent
{\bf Proof:} The diagram
\[
\xymatrix{
I \ar[r]^{\delta_0 \epsilon_0} \ar[d]^a & K \ar[d]^{c(b_1^{-1}b_0)^{\delta_0}} \\
I \ar[r]^{\delta_1 \epsilon_1} & K }
\]
commutes, since 
\[
c \delta_0 b_1^{-1} b_0 \delta_0^{-1} \cdot \delta_0 \epsilon_0 = c \delta_0 b_1^{-1} \cdot \epsilon_1 a
= \delta_1 b_1 \cdot b_1^{-1} \epsilon_1 a = \delta_1 \epsilon_1 a \;.
\] 
To check associativity, let $(a,b_0):\epsilon_0 \to \epsilon_1 \in \HEmb(I,J), \; (b_1,c_0): \delta_0 \to
\delta_1 \in \HEmb(J,K)$, and $(c_1,d): \eta_0 \to \eta_1 \in \HEmb(K,L)$; then 
\[
(a,b_0),((b_1,c_0),(c_1,d)) \mapsto ((a,b_0),(b_1,d(c_1^{-1}c_0)^{\eta_0})) \mapsto 
(a,s(c_1^{-1}c_0)^{\eta_0}(b_1^{-1}b_0)^{\eta_0 \delta_0}) \;,
\]
while 
\[
((a,b_0),(b_1,c_0)),(c_1,d) \mapsto ((a,c_0(b_1^{-1}b_0)^{\delta_0}),(c_1,d)) \mapsto 
(a,d(c_1^{-1}c_0(b_1^{-1}b_0)^{\delta_0})^{\eta_0}) \;. 
\]
$\Box$ \bigskip
 
\noindent
{\bf 2.4 Definition} As the categories $\HEmb$ are groupoids, we can define a category $\bm_\bc$ with intervals $I,J,\dots$ as
objects, and
\[
{\rm Mor}_\bm(I,J) = \pi_0\HEmb(I,J) \;;
\]
roughly speaking, it is a category of one-manifolds with boundary conditions. Note that the automorphism group of 
$I$ in $\bm$ is the quotient group $\diff(I)/\diff_0(I)$. \bigskip

\section {Conformal nets} \bigskip

\noindent
{\bf 3.1} Following Bartels, Douglas, and Henriques [[1,4]; see also [2,5,6]], a conformal net is a kind of cosheaf 
of von Neumann algebras on a category of suitably oriented compact one-dimensional Riemannian manifolds and smooth 
embeddings. This note is concerned with a weaker notion of a (continuous) functor $\fA$ on such a category, 
taking values in a category of topologized algebras, with one bit of extra structure: suppose that \medskip

\noindent
$\; \bullet \:$ to any diffeomorphism $a$ of $I$ which is supported on the interior (ie, which leaves a neighborhood 
of $\partial I$ pointwise invariant), there is an element $\fa(a) \in \fA(I)$ such that $\forall x \in \fA(I)$,
\[
\fA(a)(x) = \sigma_{\fa(a)}(x) \;,
\]
and \medskip

\noindent
$\; \bullet \;$ if $a_0,a_1$ are two such interior automorphisms of $I$, then 
\[
\fa(a_0 \circ a_1) = \fa(a_1) \circ \fa(a_0) \;.
\]
(The order reversal is intentional). \bigskip

\noindent
{\bf 3.2 Definition} I'll call such a functor a weak quantization for one-manifolds.
The condition above is stronger than the analogous (sixth) condition considered in [1 \S 3.7], but it is satisfied in 
the example below. \bigskip

\noindent
{\bf Proposition:} Such a weak quantization defines a (strict) two-functor
\[
\fA :\Bm_\bc \to \BAlg_\out
\]
which descends to a plain vanilla functor
\[
\fA_0 : \bm_\bc \to \Alg_\out \;.
\]
of ordinary categories.  \bigskip

\noindent
{\bf Proof:} To simplify notation, let $\fA(\delta) = D$; and if $a,b,\dots$
are diffeomorphisms supported on the interior of their domains, let $\fa(a) = \xa, 
\; \fa(b) = \xb$, etc.\bigskip

\noindent
The assertion amounts to checking the commutativity of the diagram
\[
\xymatrix{
\HEmb(I,J) \times \HEmb(J,K) \ar[d] \ar[r] & \HEmb(I,K) \ar[d] \\
\HHom(\fA(I),\fA(J)) \times \HHom(\fA(J),\fA(K)) \ar[r] & \HHom(\fA(I),\fA(K)) \;,}
\]
ie (using the notation above) that
\[
\xymatrix{
(a,b_0),(b_1,c) \ar[r] \ar[d] & (a,c(b_1^{-1}b_0)^{\delta_0}) \ar[d] \\
(\xa,\xb_0),(\xb_1,\xc) \ar[r] & \xc D_1(\xb_1^{-1}\xb_0) }
\]
commutes: in other words, that
\[
\fa(c(b_1^{-1}b_0)^{\delta_0}) = D_0(\xb_1^{-1}\xb_0) \xc
\]
(using the fact, from \S 1.0, that $D_1(\xb_1^{-1}\xb_0)\xc^{-1} = \xc^{-1}D_0(\xb_1^{-1} \xb_0)) \;. \; \; \; \; \; \Box$
\bigskip

\noindent
{\bf 3.3} The group $\PGl_2(\R)$ acts on the real projective line $P_1(\R)$, and the noncompact torus
\[
t \mapsto 
\left[\begin{array}{cc}
      \cosh t & \sinh t \\
      \sinh t & \cosh t
      \end{array}\right] 
\]
of Lorentz rotations preserves the interval $\xI = [-1,+1]$, defining a homomorphism
\[
\R \to \diff(\xI)/\diff_0(\xI) \;.
\]
[I am indebted to Andr\'e Henriques for pointing out the interest (and accessibility!) of this
quotient.] \bigskip

\noindent
A weak quantization $\fA$ for one-manifolds, in the sense above, thus defines a homomorphism
\[
\R \to \out(\fA(\xI)) \;.
\]
On the other hand, Connes [3] exhibits a {\bf canonical} homomorphism 
\[
\R \to \out(\fA)
\]
to the group of outer automorphisms of {\bf any} von Neumann algebra $\fA$ (and has suggested
that it be regarded as the flow defined by a kind of intrinsic time). \bigskip

\noindent
Wassermann, for example [8 \S 15] has shown that the free fermion functor $\fF$ (defined by the Fock representation
of the Clifford algebra on the $L^2$ functions on a metrized interval, cf. also [1 \S 4.1, 7 \S 4.3.5]) 
assigns a type III von Neumann algebra to a compact interval, and that it defines a conformal net for 
which the diagram
\[
\xymatrix{
\R \ar[d] \ar[r]^= & \R \ar[d] \\
\diff(I)/\diff_0(I) \ar[r] & \out(\fF(I)) }
\]
commutes.

\newpage

\noindent
{\bf 3.4} This seems to me an intriguing fact: it asserts that for the free fermion functor, Connes' intrinsic flow
agrees with the natural flow defined by the geometry of the projective line, thus providing one of the few
examples in mathematics in which Lorentz geometry appears naturally. \bigskip

\noindent
{\bf Acknowledgements} This work was supported by the NSF; it grew out of discussions at a June 2009 AIM workshop
in Palo Alto. I'd like to thank Hisham Sati for organizing that meeting, and its participants for their
interest. Andr\'e Henriques and Chris Douglas get special thanks for generously sharing their time
and ideas. \bigskip

\bibliographystyle{amsplain}

\end{document}